CME –Associated Energetic Ions at 0.23 AU — Consideration of the Auroral Pressure Cooker Mechanism Operating in the Low Corona as a Possible Energization Process

Short Title: Acceleration of CME ions by auroral process

D. G. Mitchell[1], J. Giacalone[2], R. C. Allen[1], M. E. Hill[1], R. L. McNutt[1], D. J. McComas[3], J. R. Szalay[3], N. A. Schwadron[4], A. P. Rouillard[5], S. B. Bale[6], C. C. Chaston[6], M. P. Pulupa[6], P. L. Whittlesey[6], J. C. Kasper[7] R. J. MacDowall[8], E. R. Christian[8], M. E. Wiedenbeck[9], W. H. Matthaeus[10]

[1] Johns Hopkins Applied Physics Laboratory, Laurel, MD 20723, USA;
    Email: don.mitchell@jhuapl.edu
[2] Lunar and Planetary Laboratory, University of Arizona, Tucson, AZ 85721, USA
[3] Department of Astrophysical Sciences, Princeton University, Princeton, NJ 08544, USA
[4] University of New Hampshire, Space Science Center, Durham, NH, 03824, USA
[5] IRAP/CNRS, Toulouse, France
[6] University of California at Berkeley, Berkeley, CA 94720, USA
[7] University of Michigan, Ann Arbor, MI, 48109, USA
[8] Goddard Space Flight Center, Greenbelt, MD 20771, USA
[9] Jet Propulsion Laboratory, California Institute of Technology, Pasadena, CA, 91109, USA
[10] University of Delaware, Newark, DE 19716, USA

Abstract:
We draw a comparison between a solar energetic particle event associated with the release of a slow coronal mass ejection close to the sun, and the energetic particle population produced in high current density field-aligned current structures associated with auroral phenomena in planetary magnetospheres. We suggest that this process is common in CME development and lift-off in the corona, and may account for the electron populations that generate Type III radio bursts, as well as for the prompt energetic ion and electron populations typically observed in interplanetary space.

1. Introduction:
Solar energetic particle events (SEPs) observed in interplanetary space have been associated both with flares and with coronal mass ejections (CMEs). Broadly speaking, the acceleration of SEP energetic particles is often associated with one or both of two processes, reconnection and shocks, with reconnection generally describing the acceleration of particles in solar flares [e.g., Zharkova et a. (2011)], and shock acceleration for the case of CME-driven shocks [Reames (1999); Melrose (2007)]. While shock acceleration has been extensively modeled and is quite well understood [Krymsky, 1977; Axford et al. (1978); Bell (1978); Blandford and Ostriker (1978); Giacalone et al. (2012); Schwadron et al. (2015)], the detailed processes by which reconnection accelerates ions and electrons to high energies (100's of keV to MeV) are rarely explicitly discussed. Rather, energetic particles are usually simply attributed to reconnection, without providing explicit processes by which the particles are accelerated (among some exceptions being Cairns et al. (2003); Drake et al. (2009); Zank et al. (2000), Guo et al. (2019), in which island collapse and Fermi-like processes are invoked). Another acceleration mechanism has been invoked in an attempt to account for events with enhanced 3He/4He ratios, involving resonant acceleration by waves generated by outward streaming beams of energetic electrons [Temerin and Roth (1992); Roth and Temerin (1997)]. This mechanism, though referred to in more recent literature (e.g., Reames (2015), has not gained broad acceptance because of perceived deficiencies in explaining the associated heavy ion enrichment in the 3He-rich events.

Electrons accelerated in these events close to the sun have been definitively associated with Type III radio bursts [e.g., Reid and Ratcliffe (2014)], in which beams of field-aligned electrons generate Langmuir waves that then mode convert to produce electromagnetic radiation at the local plasma frequency (fundamental) and at the second harmonic. Their time-frequency structure is directly linked with the traversal of the electron beam on open field lines through the decreasing plasma density from the chromosphere/low corona into the high corona and on into interplanetary space. A variant of these is attributed to field-aligned electrons trapped on closed loops, in which case the time-frequency structure folds back upon itself as the electron beam first gains altitude, then returns to the lower corona. A second type of radio emission, Type II, is associated with particularly large and fast CMEs, and is attributed to electron acceleration in shock waves produced as these fast CMEs propagate through the corona and solar wind at velocities faster than the local fast magnetosonic speed. Such shocks also accelerate some solar wind ion populations to high energies, typically over a broader set of coronal foot point connections than is the case for the sources of the Type III event electrons. In this paper, we will focus on the particle population associated with Type III events and CMEs, not the shock-associated particles (although the former may sometimes provide seed particles for the latter).

Energetic ions are also often produced during the formation of relatively slow CMEs, and these can also populate open field lines and be detected at least as far out as 1 AU. Their composition is somewhat variable, but generally consistent with coronal composition, presumably reflective of their source plasma.

## 2. CME event of November 11, 2018

On November 11, 2018, at a distance of about 0.23 AU from the sun, several in the suite of instruments aboard the Parker Solar Probe (PSP) recorded data on the solar wind plasma density, velocity and electron heat flux (SWEAP; Kasper et al., 2016), magnetic field and radio emissions (FIELDS; Bale et al., 2016), and energetic particles (ISOIS; McComas et al., 2016) that characterized a set of phenomena associated with the lift-off and subsequent propagation to, and measurement by, PSP of a coronal mass ejection (CME) at 0.25 AU. ISOIS is comprised of two sensors, EPI-Lo covering ion energies from ~30 keV to 100 MeV, and EPI-Hi covering ion energies from ~2 MeV/nucleon to galactic cosmic ray energies. The basic characteristics of this event (detected only at the lower energies measured by the EPI-Lo instrument, but not by EPI-Hi) were described by McComas et al., (2019). The event is also described in considerable detail in a companion paper [Giacaloni et al. (2019)] in this issue. Giacalone et al. (2019) discuss the time sequence, along with a proposed mechanism of acceleration of the energetic particles by a traveling pressure enhancement of a seed population in the solar wind plasma ahead of the rather slowly propagating CME. Giacalone et al. (2019) also go into the effects of transport on the timing and anisotropy of the energetic particle observations, and we will not repeat that analysis here. Rather, we present complementary aspects of the data, emphasizing the composition and energy range of the energetic particles. We argue, based in part on the conclusion that it was unlikely that there was a well-positioned shock associated with this event, (see Rouillard et al. (2019), this issue), that the event may be a good candidate for being produced by current-driven processes [as also suggested for downward accelerated electrons by Cairns et al. (2003), and Haerendel (2017)], possibly associated with the reconnection that must take place during and after the release of a CME, or simply driven by the relaxation of the region to a less stressed configuration of the corona after the release of the stressed fields and large mass carried away by the CME [e.g., Chen (2011)].

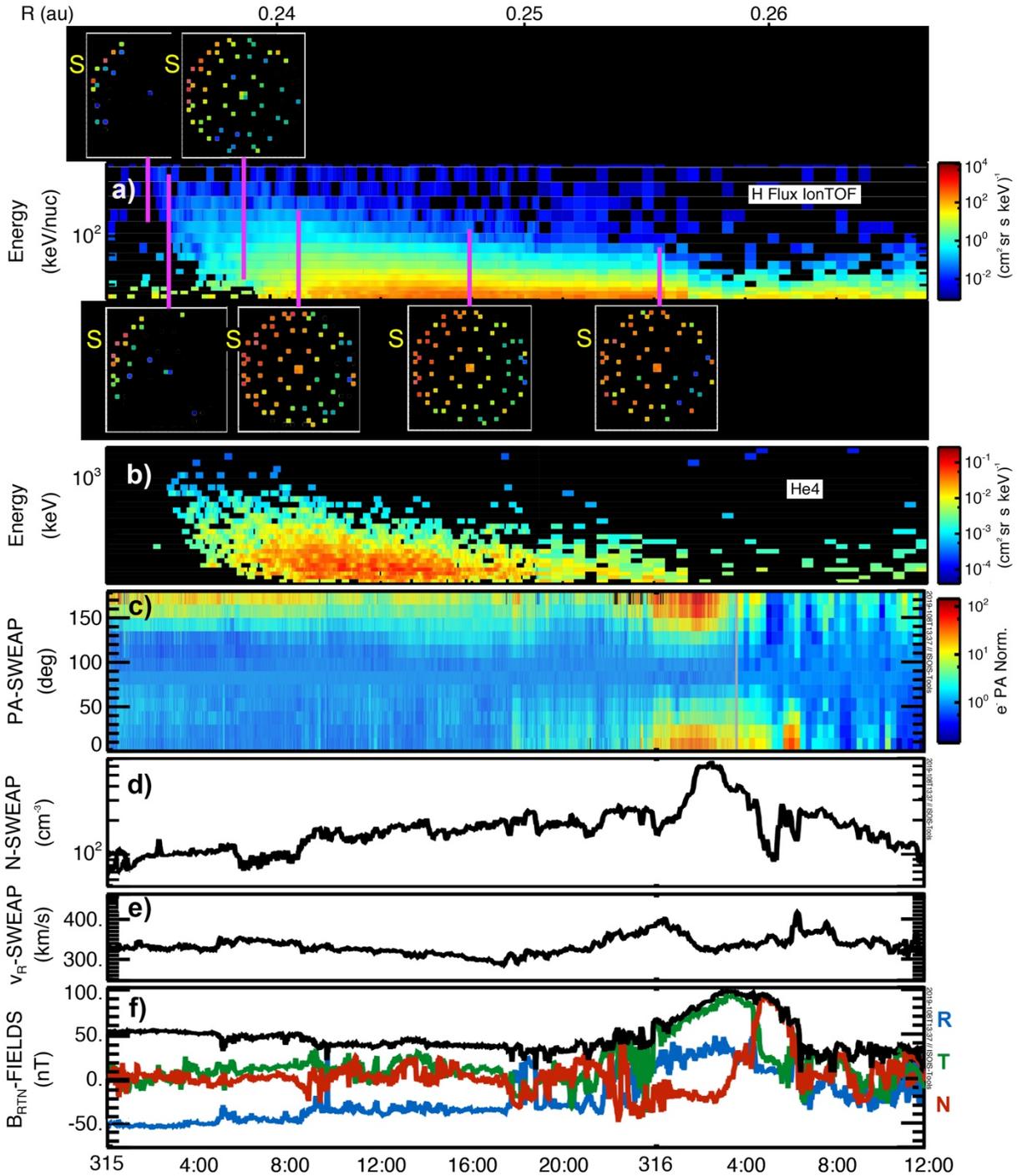

*Figure 1. From the top, a) (with accompanying anisotropy diagrams), energetic ions as measured by the EPI-Lo time of flight (TOF) -only technique; b) energetic helium spectrogram, showing energies exceeding 1 MeV at the leading edge, about 1 day before the arrival of the CME at PSP; c) electron heat*

*flux, showing outward field-aligned streaming electron prior to the CME, becoming bi-directional inside the CME; d) solar wind density, with a strong enhancement inside the CME structure; e) solar wind velocity, showing just a modest enhancement in the flow velocity in the plasma leading the CME; (f), the three components (R radial, N north-south, and T east-west) plus the magnitude of the solar wind magnetic field with the signature of a CME passing over PSP at the beginning of DOY 315.*

In Figure 1 we present an overview of the event, with measurements by the aforementioned instruments both for context, and to provide clues to the energetic particle acceleration process. Panel a) shows the energetic ions as measured using only their TOF (which permits us to measure particle velocities below the minimum needed to measure the particle energy in the EPI-Lo solid state detectors; the vast majority of these are protons, so we assume mass 1 in deriving the ion energy from the measured velocity). We also include small anisotropy plots, that show the relative intensity seen in each of the 80 EPI-Lo apertures that cover ~ ½ the sky in angular coverage. Each anisotropy plot represents an hour average centered at the time in the ion spectrogram indicated by the pink vertical bar leading to the plot. Similar plots can be found in Giacalone et al. (2019), (this issue), where they are discussed in more detail and where inferences regarding the propagation of the particles from the low corona to PSP are drawn. The spectrogram in panel a) shows that the protons reach energies of at least several hundred keV at the leading edge of the event. In b) the helium reaches energies over 1 MeV at the leading edge. The context measurements in panels c) through f) will not be discussed in detail here, as they are well covered for this same event in other papers in this issue (e.g., Korreck et al. (2019)). The primary points that these context measurements illustrate are that 1) this is a relatively slow CME with no indication of a local shock; 2) the magnetic field is mostly radial prior to the arrival of the CME and the plasma just ahead of the CME; 3) the energetic particles are detected primarily on the open magnetic field prior to the CME, and drop to much lower intensities and lower maximum energies inside the CME structure.

The CME signatures are quite interesting, exhibiting a change of magnetic polarity of the component N of magnetic field at 4:00 on day 316 possibly indicating passing through X-null point of a reconnecting current sheet. Furthermore, there is a strong increase of T-magnetic component about the same time indicating a noticeable guiding field in this current sheet. These features are important to understanding the CME itself, and may be related to processes associated with particle acceleration. However, as indicated by the SWEAP pitch angle information in panel C, the field topology of the CME structure is mostly closed, and not a good candidate for the source of the energetic ions observed on open field lines almost a day earlier, the subject of our analysis in this paper.

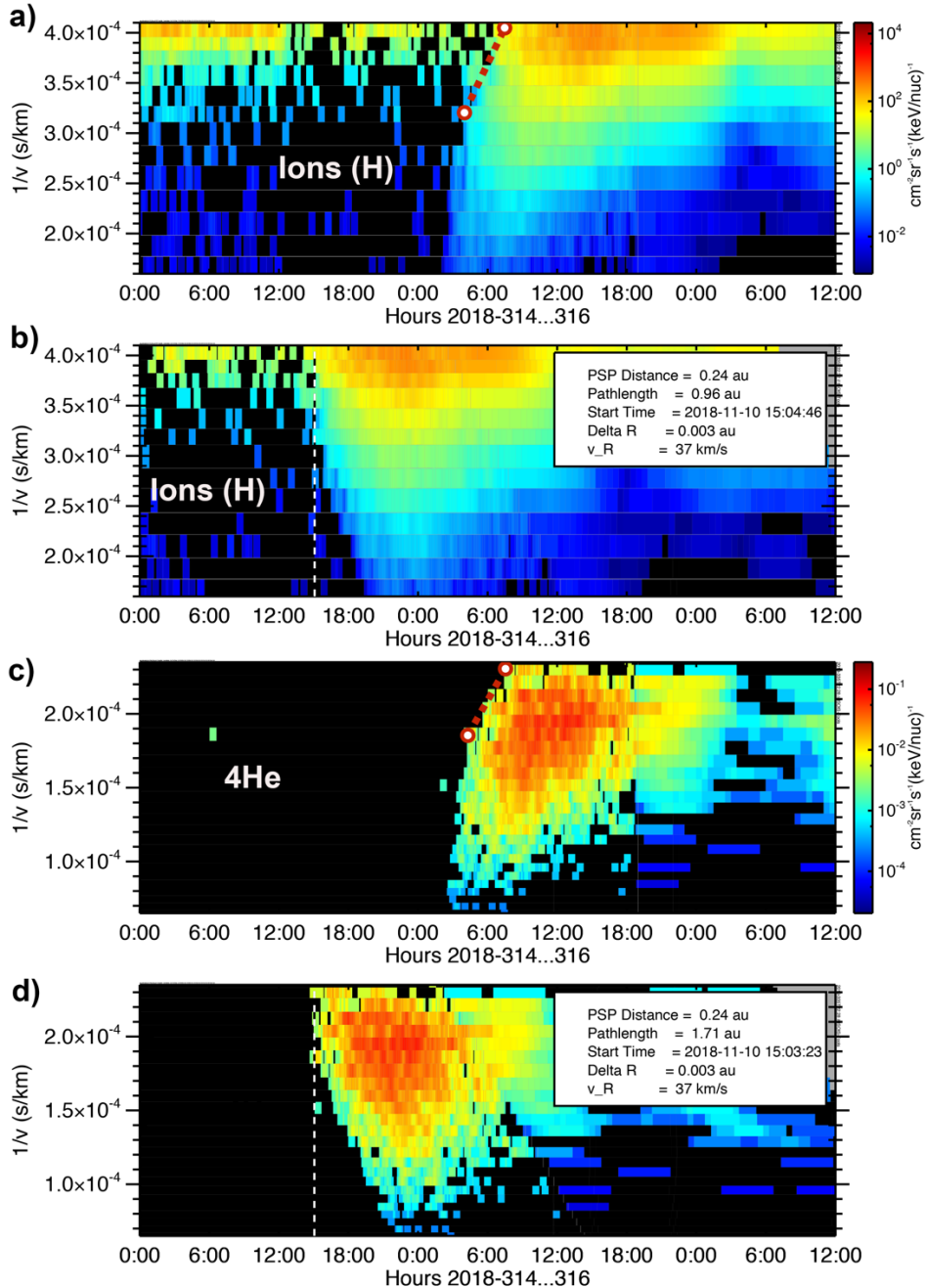

Figure 2. a) Ion (assumed hydrogen) intensity as a function of velocity on an inverted velocity scale. A straight line is fit to the leading edge of the

*low energy portion of the event, from the slowest to intermediate velocity particles with intensities persistently above background levels. b) same data, but shifted in time by the velocity dependent propagation time from the low coronal to PSP obtained from the fit in a). The pathlength derived from this fit for the particles is found to be greater than the radial distance of the spacecraft from the sun, implying considerable scattering in pitch angle along the path. c) 4He intensity as a function of velocity on an inverted velocity scale. As in a), a fit to the leading edge of the low energy portion of the event is made. d) same 4He data, shifted in time by the velocity dependent propagation time from the low coronal to PSP obtained from the fit in c). The pathlength derived from this fit is also greater than the radial distance of the spacecraft from the sun, again consistent with scattering.*

In Figure 2 we show an analysis of the velocity dispersion for ions (hydrogen) and helium measured just at the leading edge of the event, where the anisotropies are most beam-like and the particles are therefore presumed to be the least scattered among those propagating from their acceleration altitude. This analysis shows that the distance traveled by these particles from their source was greater than the 0.24 AU distance from PSP to the sun. In Figures 2b and 2d, the data are shifted in time by the fit propagation time of the leading edge, to represent the time profile of their acceleration at their source in the low corona. For both the ions (hydrogen) and the helium, the derived source injection time profile suggests the energization of lower energies early in the event, with higher energies adding in until the peak energies for each species are obtained about 8 hours after the initial injection, then slowly tapering again to only relatively lower energies over the ensuing 12 to 14 hours. How much of this tapering is propagation related (e.g., from diffusive transport) and how much is intrinsic to the source time-profile is not possible to determine from this technique alone. To the extent that this treatment can reverse the distortions of the source time-profile induced by propagation, the data are consistent with a gradual build up in the energies that can be produced through the acceleration process, rather than an impulsive injection at the full range of observed energies. The arrival of the highest energies before the lower at PSP is, according to this treatment, a consequence of propagation, and not a consequence of their acceleration time profile. This analysis indicates that the acceleration process began at about 1500 UT on November 10. At this time, the CME had not yet been imaged in the STEREO

Cor2 coronagraph field of view (it appeared about 4 hours later) and so remained below ~3Rs.

Had we used the leading edge of the high energy portion of the observed event to fit the propagation delays for all the particles, the event would have fit a source acceleration profile more like that obtained in Giacalone et al. (2019), that is, an injection over all energies together at a common injection time, early on November 11. Which approach is more representative of the true acceleration time profile is not important to the mechanism we will discuss below, but we include this alternate analysis to point out the ambiguity in this approach to trying to obtain the source acceleration time profile, and to point out that one's choice of which portion (in energy) of the leading edge to fit significantly affects the stage in the evolution of the CME to which one ascribes the acceleration process. If the approach employed here is applicable, either mechanism would have to be able to reproduce the gradual rise in peak energy.

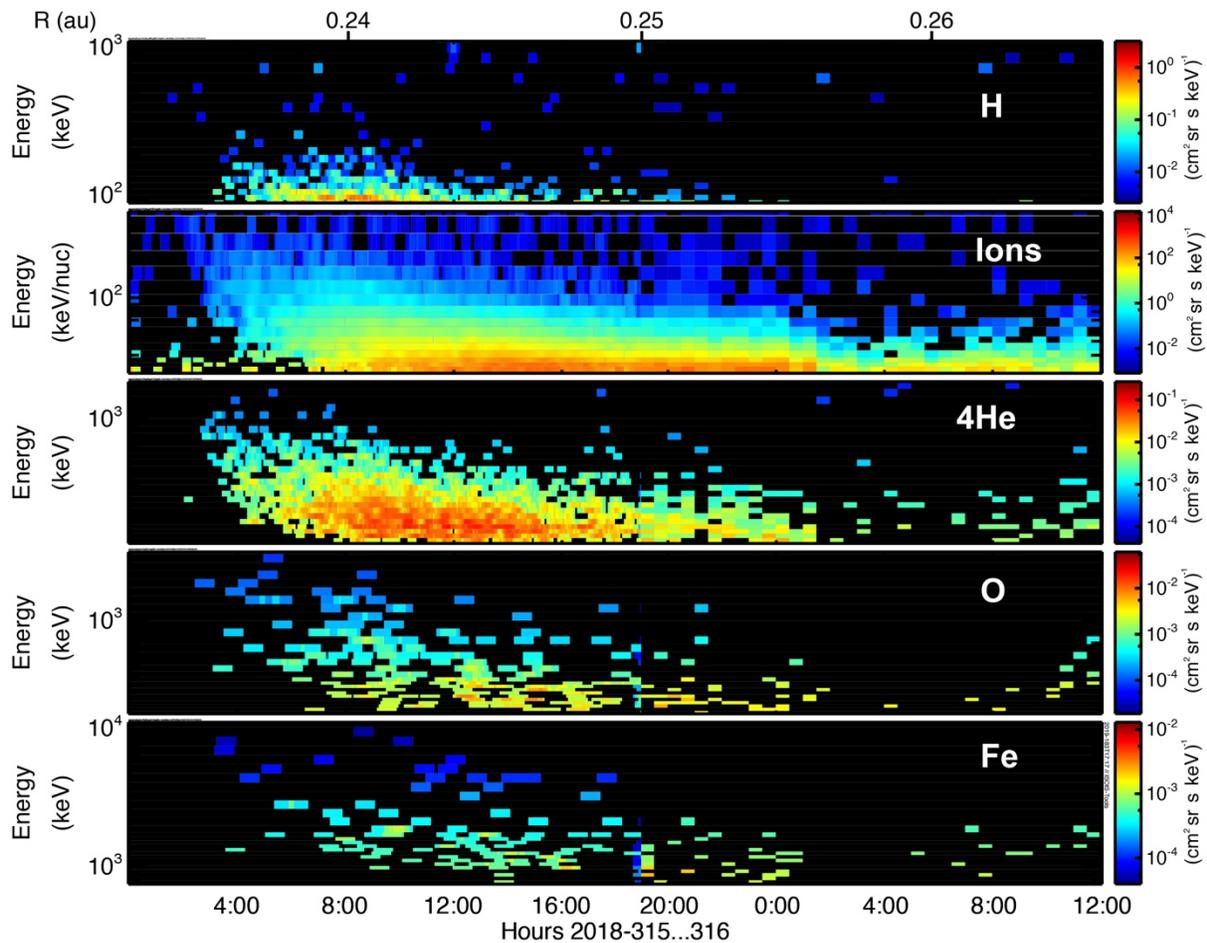

*Figure 3. Composition of the energetic particles in the November 11 SEP. Each panel is an ion spectrogram, intensity as a function of particle energy plotted against hour on November 11 and 12, 2018. From the top, the spectrograms are for protons; ions interpreted as protons measured in velocity only; 4He; O; and Fe. Note that the energy scale is in keV, and that the scale varies for each ion species.*

We now focus on the composition of the ions accelerated in this event, and the peak energies reached by each species. The event included ions typical of coronal plasma, and typical of SEP events. Those signals well above the EPI-Lo background for this very modest event included H, 4He, O, and Fe (Figure 3). There is a hint of 3He measured, but with statistics insufficient to cite a 3He/4He ratio. The highest energies for which counts above background were measured increased as a function of particle mass (and probably with increasing charge state, although we have no measure of charge state with this instrumentation), with protons seen (in the TOF-only ion data) up to ~400 keV, 4He to ~1 MeV, O to ~2 - 3 MeV, and Fe to ~8 MeV.

### 3. Discussion:

The composition is suggestive of an acceleration process operative in the corona at an altitude where the plasma is transitioning from collisional to non-collisional that depends on either charge state or mass, or both. Since many acceleration processes boil down at the most fundamental level to energy gained as a given charge passes through a potential difference, charge state is a natural parameter to examine for ordering the energy gained.

Typical charge states for the measured constituents for the solar wind are H = +1, 4He = +2, O = +6/7, Fe = +10/11/12. For impulsive SEPs the heavier ions tend toward higher charge states characteristic of higher temperatures in the source plasma (up to ~20 for iron) [Moebius et al. (1999); Popecki et al. (2002); Desai and Giacalone (2016)]. In the following discussion we use the solar wind charge states (which are also close to those for energetic ions in gradual solar events), since there was no flare observed in the region where this CME originated. If we use 4He as the reference (since our signal to noise is better for 4He than it is for the other ions), and we use 1 MeV for the peak 4He energy, then we should expect ~500 keV for peak H energy, 3 MeV for peak O energy, and 5 to 6 MeV for peak Fe energy. The observed peak values fall very close these ranges. In the case of TOFxE H, the EPI-Lo efficiency drops steeply for H above about 100 keV, so the lower maximum energy observed there may simply be a consequence of the signal being well below the one count level. However in the TOF-only ions, which should be dominated by protons, enhanced intensity was seen up to 400 keV. Thus, these observations are consistent with a process that involves charge dependent energization. The slightly higher than predicted energy reached by Fe may indicate slightly higher charge state for iron, ~ +16.

It has long been accepted that more than one mechanism must operate to energize ions in SEP events, primary among them being shock acceleration and reconnection, as well as wave-particle acceleration. Giacalone et al. (2019) favor a shock acceleration type of process for this event. SEPs have typically been viewed as candidates for shock acceleration, and for many the spectra, propagation, and timing have been successfully associated with that mechanism. In the case of this CME, Rouillard et al. (2019) (this issue) estimate that the CME regions magnetically connected to PSP do not appear to develop a shock during the entire CME propagation. Their calculations do suggest that a sub-critical shock (Ma<3) could have formed at the surface of the (idealized) CME flux rope at other longitudes, induced by the 3-D expansion of the CME in the frame of the solar wind. Particles may have been accelerated at that weak shock location (as posited in Giacalone et al., (2019).

However it is not obvious that particles accelerated at a weak shock in that location would have found their way to the magnetic fields connected to PSP. Our analysis of the timing of the dispersion observed indicates that the energization of the highest energy ions must have begun about 1500 UT on November 10 (day 314), well before the time of the appearance of the CME in the STEREO A/COR2 instrument at about 2000 UT on day 314, according to Giacalone et al., (2019). The high anisotropy and the dispersion analysis (Figure 2) indicating a source near the sun suggest that the ions had ready access to topologically open magnetic field (connected with PSP) at the time of their energization, which was when the CME was still relatively close to the sun (below ~3 to 4 solar radii, Figure 1 of Giacalone et al. (2019)).

As a CME forms and lifts off from the corona, topological changes are taking place in the coronal magnetic field surrounding it. The CME itself is often a magnetic configuration with considerable shear indicating the presence of strong currents, and reconnection is likely taking place in the thin current sheet behind the plasmoid or flux rope that forms the core of the CME, as well as possibly between the legs of the closed field draped over the flux rope and either open or closed field lines adjacent to those closed lines that are still rooted in the corona and chromosphere (e.g., Crooker et al., 2002). These processes should also involve strong currents, both across the reconnection region, and diverted along the magnetic field where they flow as field-aligned currents back into the low corona, chromosphere and photosphere.

It is commonly assumed that field-aligned currents in the low corona can be approximated by vertical photospheric currents $Jz$ (which can be calculated from vector magnetograms--see the analysis in Janvier et al. (2014) to illustrate this approach—their Figure 2 and the discussion of vertical currents). In the force-free approximation for the low $\beta$ corona, during times of low activity these field-aligned currents follow closed loops and re-enter the photosphere at the loop foot-points. Above the Alfven critical point (which is really distributed over an altitude range, not a point—more on this below), the plasma transitions to $\beta > 1$, and the force-free approximation no longer holds. CME eruptions usually take place in regions with high magnetic field strength, considerable shear, and strong photospheric vertical currents [e.g., Chen (2011); Janvier et al. (2014); Georgoulis (2018); Georgoulis et al., (2019); Vourlidas and Webb (2018)]. This photospheric current pattern is not greatly altered during the CME growth and lift-off [(Janvier et al. (2104)], so these currents (those that map to the CME structure including the core flux rope and the reconnection current sheet that forms beneath it, and those connecting with the over-arching loops that may be carried outward with the CME front) continue to flow as the CME lifts off.

Additional currents may also be generated by the CME lateral expansion, as the structure moves through the corona and forces aside adjacent pre-existing loops and open field (generating J x B currents in the process, e.g. Delannée et al. (2008)). In this process, some of the strong field-aligned currents that initially flowed entirely within the low corona, where the plasma electrons could easily provide that current, close through increasingly high altitude loops and/or open field, where the plasma density falls off quickly and quasi-neutrality may no longer support the parallel electron current without the development of a parallel potential, analogous to the auroral case.

A third source of field-aligned currents in the corona and transition to the solar wind are the field-aligned component of the currents that are required to support corotation of the plasma. PSP results (Kasper et al. (2019) show that corotation of the coronal/solar wind plasma continues much farther from the sun than pre-PSP modeling predicted. The solar wind plasma continues to include a corotation component as far out as 40 Rs. As has been shown repeatedly for planetary magnetospheres, the torques required to keep the radially transported plasma corotating or sub-corotating are transmitted by field-aligned currents from the foot-points of the magnetic field (in the ionosphere for Jupiter and Saturn, or in the photosphere for the sun) to the rotating plasma. This process transfers angular momentum from the sun or planet to the plasma (Hill 1979; Bagenal 2007) and the field-aligned currents are necessary for that transfer. So the entire corona will have a continuously driven set of field-aligned currents, whose magnitude will change depending on the rate at which angular momentum is being transferred. Conditions that would lead to enhancement of the currents that make up this system would include an increase of either the mass of the plasma being accelerated radially, or the speed at which it is moving radially through the region that corotates or sub-corotates. Cleary the injection of a CME would involve an increase in the mass of the plasma, and frequently an increase in the speed as well. Such conditions would lead then not only to the creation of the current system discussed in Delannée et al. (2008), but also to enhancements in the corotation enforcement field-aligned currents associated with this transfer of angular momentum into the radially accelerating solar wind.

Thus there are at least three reasons that one should expect field-aligned currents to be enhanced in association with the injection of a CME, which is just what would be required for the pressure cooker conditions to occur.

In planetary magnetospheres, strong field-aligned coupling currents are inescapable features both of sheared flows (for example in the cusp [e.g., Gorney et al., (1985)], low latitude boundary layer or in drag-induced subcorotation flows [e.g., Cowley et al., (2008)]), and of reconnection (substorms, plasmoid/flux-rope production, e.g. Artemyev et al., (2018)). These currents often drive the development of field-aligned potentials in the ionosphere-to-magnetosphere transition whereby the plasma electrons are accelerated in order to maintain the current with sufficiently few electrons that charge neutrality of the plasma is not violated. The resultant anisoptropic energetic electrons drive instabilities and result in the growth of broadband extremely low frequency (BBELF) waves perpendicular to the magnetic field. These broadband waves cover a frequency range where they interact efficiently with the plasma ions by ion cyclotron resonance with the various species [e.g., Lynch et al., (2002); Jasperse (1998); Carlson et al., (1998)], heating them in the direction perpendicular to the magnetic field. In downward electric current regions, the electric potentials accelerate the electrons upward along the magnetic field, while the heated ions, which would otherwise quickly escape the region by virtue of the mirror force in the diverging magnetic configuration, are confined by the same potential so that they remain trapped in the wave field generated by the electrons and gain even more energy. Eventually, the ions gain sufficient energy perpendicular to the magnetic field that the mirror force becomes great enough to push the ions through the confining electric field force, and they then escape upward as energetic ion conics. This process can result in ions with considerably higher total energy than the potential drop confining them represents, since the energy parallel to the magnetic field that is required to overcome the electric potential will be much less than the perpendicular energy of the nearly 90 degree pitch angle ions. As they escape the wave heating region, they travel out into weaker and weaker magnetic field where their conic angles close and become nearly parallel to the magnetic field, so that at large altitudes they are seen to be streaming close to field-aligned.

In auroral processes at Earth, this mechanism results in perpendicularly heated ions with typical energies of a few keV [Gorney et al. (1985), who coined the descriptor "Pressure Cooker" for this process; Lynch et al., (2002); Jasperse (1998); Ergun et al. (1998); Carlson et al. (1998); Klumpar et al., (1990); Marklund et al., (2001)]. At Saturn, Mitchell et al. (2009) have shown that the same process results in much higher ion energies (~200 keV) while at Jupiter, Clark et al. (2017) show ion energies exceeding 1 MeV. Mitchell et al. (2009) suggested that the energy reached by the particles accelerated through this process scales with the system size, and the Jovian results indicate that magnetic field strength at auroral altitudes also results in scaling the ion energy higher.

Although the concept of the "Pressure Cooker" was originally investigated in Gorney et al. (1985) under the scenario of an electromagnetic ion cyclotron (EMIC) wave field providing the energization mechanism for the ion energization, most of the subsequent reference to the concept was applied in scenarios for which the wave field was observed to be broad band extremely low frequency electrostatic waves (BBELF waves). It was generally found that the more intense events exhibited BBELF waves, with EMIC wave events in the minority. Whereas the EMIC-associated events favored He+ heating, the BBELF waves heated all charged ion species, presumably because sufficient wave power was available at each species' gyration frequency provided that the ions were confined to that region long enough. In all of these cases the term "Pressure Cooker" was applied, based on the presence of the parallel electric potential that confined the ions to the wave region where they would continue to gain perpendicular energy until the mirror force was sufficient to overcome the electric potential. Jasperse (1998) published a kinetic model that reproduced the Gorney et al. (1985) observations quite well. Although in Jasperse (1998) the BBELF waves were assumed to be present, they were not produced within the model. This was not a particular liability, as the BBELFs were measured and known to be present, as was typical of such events [e.g., Lynch et al., (2002)].

In the case of the solar corona, the parameters are somewhat different. Temerin and Roth (1992) explored the range of parameters in auroral plasmas compared with possible flare plasma conditions (their Table 1), and found that the auroral mechanism for (in their case EMIC) wave generation and ion resonant heating would likely work in the corona as well. However, nothing in their treatment precludes the development of BBELF waves, extending the ion heating to all species present. The coronal temperatures are much higher than those in ionospheres, producing much higher scale heights in the plasma at the altitude where this process would be expected to take place (above the collisional region, so that the field-aligned potentials freely accelerate the electrons). This may result in different electric potential structures, which could either increase or decrease the total potential drop, depending on the requirements for current density and the supply of electrons from the quasi-neutral plasma. However, it is quite likely that such structures can be generated, and if they are, then ion heating and escape is inevitable, in which case, even in the absence of a shock to accelerate the ions, energetic ions would still be produced.

Charge state dependence would be a natural consequence of this scenario. In the planetary auroral situation, where all of the ions are typically singly ionized, the various species all reach approximately the same peak energy, inconsistent with a process dependent on mass, but consistent with a process dependent on confinement by a potential. As in the auroral case, for a similar process in the corona the mirror energy required to overcome the confining potential should be proportional to the ion net charge, so that from a given altitude in what is undoubtedly an extended region of potential drop any higher charge state ions present would need to reach a higher perpendicular energy to overcome that potential, again proportional to the net charge. This is consistent with the observed energy dependence of the ions in this event.

In the auroral case, the electric field is typically distributed over a considerable distance in altitude, as a series of double layers with small potential drops across each individual double layer. As discussed above, in the auroral case direct measurement of the waves show a broadband distribution of waves, presumably able to accelerate ions with a broad range of charge/mass ratios as sufficient wave power would be present at each species' resonant frequency. This multiple-species response to the broadband waves is confirmed in that for Earth, the ion energization is independent of mass [Auroral Plasma Physics (2012), pp 125; Lund et al., (2001); and Lynch et al. (2002), who also published wave spectra confirming the adequacy of the BBELF power], for Saturn the primary ion constituents are $H^+$ and $H_3^+$ (i.e., ionospheric),[Mitchell et al. (2009)] and at Jupiter H+, O+, and S+ are accelerated [Clark et al. (2017)]. The broadband wave field is likewise distributed in altitude, so that ions are heated over a broad altitude range. In order to be observed, those at the lowest altitude in the potential structure must achieve the highest energies, since they must overcome the greatest total potential barrier. The electrons accelerated upward by the electric potential account for the (downward) electric current.

As Temerin and Roth (1992) suggested, it is quite likely that coronal conditions can support the same physics that takes place in the auroral downward current regions. They explored the effects of waves generated by streaming electrons on the ambient ion population, and concluded that ion heating should take place in the corona based on a similar mechanism. However, their focus was on 3He enhancement, and they concentrated on resonant acceleration of 3He by EMIC waves. Furthermore, they did not consider the effect of the parallel electric field on confining the ions in the accelerating wave field. With these additional elements, and the broadband nature of the wave field, we find this (the pressure cooker mechanism) to be an attractive candidate for at least some of the SEP events produced during CME lift off, as well as possibly for flares (although in our case, no flare was observed). While we have no direct measurement of the BBELF waves, and for this event no measurement of the field-aligned electron beams, neither of these is surprising. We can't measure the BBELFs since we do not have measurements in the acceleration region and electrostatic waves will not propagate to PSP. It is not clear what behavior should be expected for the field-aligned electrons that should be produced in the event. At Earth, Carlson et al. (1998) showed that the peak electron energy lay just below the minimum ion energy in the observed ion conics, and the ion energies extended well above the electron peak energy. On the other hand, at Saturn Mitchell et al. (2009) observed electron peak energies that exceeded the highest detected ion conic energies. These differences may reflect different parametric limits for the two locations. If the wave field generated by the electrons can efficiently accelerate the ions to energies much higher than the total potential drop that accelerates the electrons, then ions throughout the extended region of the potential structure should gain sufficient energy to overcome the potential, and the Earth-like scenario should play out. At Saturn, although the largest total potential drop as indicated by the peak electron energies is several hundred keV, the waves excited by the electrons appear only to efficiently accelerate ions to a few hundred keV/charge perpendicular energy, so only those ions accelerated in the upper reaches of the potential structure have sufficient mirror force to escape; those deeper in the structure would be heated, but do not reach energies sufficient to escape. So for the Saturn case the relationship between the maximum electron energy and the minimum ion energy as seen for the Earth case no longer holds. A factor that could explain this limit on the peak ion energies attained at Saturn is the coherence length of the BBELF waves perpendicular to the magnetic field. Once the gyroradius of an ion exceeds the coherence length of the waves, the heating efficiency would drop quickly as the driving fields would no longer be resonant over the ion gyration diameter.

For the event discussed here, no electrons were observed above the detector backgrounds. There are several possibilities for the absence of the electrons. 1) The electrons may have propagated beyond the spacecraft before PSP was magnetically connected with the coronal region from which the ions propagated; or, 2) the electrons may have been present, but only at energies below the minimum detection energy for the EPI-Lo SSDs (~40 keV); or, 3) energetic electrons may have been present only near zero degrees pitch angle, and the magnetic field may not have fallen within the FOV of any of the EPI-Lo apertures; or 4) since this is a weak event, even in the observed ions, and as our electron measurement is a single parameter measurement, the background for electrons is orders of magnitude higher than the background for the ion measurement and so electrons may be present, but not above the detector background. However, it should also be stated that there were no Type III bursts detected above the FIELDS instrument background, so any electron beams that may have been generated were likely very weak, and so unlikely to be detectable by EPI-Lo.

Investigation of these parameters in the corona remains for future work. Since Type III bursts have been associated with electrons in the energy range of ~8 keV, it would be quite possible for Type III bursts to be generated without any electrons being detected above the 40 keV EPI-Lo energy threshold. If in this case the pressure cooker mechanism is responsible for the electron acceleration only to energies below the EPI-Lo 40 keV threshold, ions accelerated to considerably higher energies than the electron characteristic energy would still be expected (as is routinely the case at Earth). Since no Type III bursts were seen in this event, it is likely that either that electrons were not accelerated to energies as high as 8 keV, or that if they were, there were not sufficient numbers to produce Type III bursts above background.

## 4. Acknowledgements:

This work was supported by the ISOIS instrument suite on NASA's Parker Solar Probe Mission, contract NNN06AA01C. The data supporting the analysis of the energetic particle event was measured by the EPI-Lo instrument of the ISOIS instrument suite, D. J. McComas Principal Investigator. The lead author would like to thank the engineering team at the Johns Hopkins Applied Physics Laboratory, in particular G. B. Andrews, L. E. Brown, S. A. Cooper, R. S. Gurnee, J. R. Hayes, R. S. Layman, K. S. Nelson, C. W. Parker, C. E. Schlemm II, M. R. Stokes, A. R. Dupont, J. C. Hutcheson, M. W. Noble, H. Seifert, K. Strohbehn, and J. D. Vandegriff.

## 5. References:

Artemyev, A. V., Pritchett, P. L., Angelopoulos, V., Zhang, X.-J., Nakamura, R., Lu, S., et al. (2018). Field-aligned currents originating from the magnetic reconnection region: Conjugate MMS-ARTEMIS observations.

Geophysical Research Letters, 45, 5836—5844. https://doi.org/10.1029/2018GL078206

Auroral Plasma Physics (2012), Götz Paschmann, Stein Haaland, Rudolf Treumann Editors, Springer Science & Business Media, Dec 6, 2012

Axford, W.I., E. Leer, and G. Skadron (1978), The acceleration of cosmic rays by shock waves, 15th ICRC, (Plovdiv), 11, 132.

Bagenal, F. (2007), The magnetosphere of Jupiter: Coupling the equator to the poles, Journal of Atmospheric and Solar-Terrestrial Physics 69, 387—402

Bale, S. D., et al. (2016), Space Sci. Rev., 204, 49-82.

Bell, A.R., The acceleration of cosmic rays in shock fronts. I , *MNRAS*, 182, 147, (1978).

Blandford, R.D., and J.P. Ostriker (1978), Particle acceleration by astrophysical shocks, *Astrophys J.*, 221, L29.

Cairns, I.H., Lobzin, V.V., Donea, A. *et al.* Low Altitude Solar Magnetic Reconnection, Type III Solar Radio Bursts, and X-ray Emissions. *Sci Rep* 8, 1676 (2018) doi:10.1038/s41598-018-19195-3

Cairns, I. H., et al. (2003), Type II Solar Radio Bursts: Theory and Spaceweather Implications, Space Science Reviews 107: 27—34

Carlson, C. W., et al. (1998), FAST observations in the downward auroral current regions: Energetic upgoing electron beams, parallel potential drops, and ion heating, Geophys. Res. Lett., 25, 2017 — 2020, doi:10.1029/98GL00851.

Chen, P. F. (2011), Coronal Mass Ejections: Models and Their Observational Basis, Living Rev. Solar Phys., 8, 1

Clark, G., et al. (2017), Observation and interpretation of energetic ion conics in Jupiter's polar magnetosphere, Geophys. Res. Lett., 44, 4419—4425, doi:10.1002/2016GL072325.

Cowley, S.W.H., et al., (2008), Auroral current systems in Saturn's magnetosphere: comparison of theoretical models with Cassini and HST observations Ann. Geophys., 26, 2613—2630

Crooker, N. U., J. T. Gosling, S. W. Kahler (2002), Reducing heliospheric magnetic flux from coronal mass ejections without disconnection, J. Geophys. Res. 107, 1028-1032, doi:10.1029/2001JA000236

Delannée, C., T. Török, G. Aulanier, J.-F. Hochedez, (2008), A New Model for Propagating Parts of EITWaves: A Current Shell in a CME, Solar Phys., 247, 123—150.

Desai, M. I., and J. Giacalone (2016), Large gradual solar energetic particle events, *Living Reviews in Solar Physics*, 1 3, article id. 3.

Drake et al. (2009), A magnetic reconnection mechanism for ion acceleration and abundance enhancements in impulsive flares, Astrophys. J., 700:L16–L20, doi:10.1088/0004-637X/700/1/L16

Ergun, R. E., et al. (1998), FAST satellite observations of electric field structures in the auroral zone, Geophys. Res. Lett., 25, 2025 – 2028, doi:10.1029/98GL00635.

Georgoulis, M. K., Nindos, A., Zhang, H., (2019) The source and engine of coronal mass ejections. Phil. Trans. R. Soc. A 377: 20180094. http://dx.doi.org/10.1098/rsta.2018.0094

Georgoulis, M. K., (2018), The Ambivalent Role of Field-Aligned Electric Currents in the Solar Atmosphere, Electric Currents in Geospace and Beyond, Geophysical Monograph 235, Chapter 22, First Edition, Edited by Andreas Keiling, Octav Marghitu, and Michael Wheatland, © 2018 American Geophysical Union, Wiley; https://agupubs.onlinelibrary.wiley.com/doi/abs/10.1002/9781119324522.ch22

Giacalone, J. (2012), Energetic Particles Associated with Strong Interplanetary Shocks, Astrophys. J., 761, 28, 2012.

Giacalone, J., D. G. Mitchell, M. E. Hill, et al. (2019), Astrophys. J., in prep. [this issue]

Gorney, D. J., Y. T. Chiu, and D. R. Croley (1985), Trapping of ion conics by downward parallel electric fields, J. Geophys. Res., 90, 4205–4210, doi:10.1029/JA090iA05p04205.

Guo, T., et al. (2019) Sci. Adv. 2019;5: eaau7004

Haerendel, G. (2017), Evidence for Field-parallel Electron Acceleration in Solar Flares, The Astrophysical Journal, 847:113, doi:10.3847/1538-4357/aa8995

Hill, T. (1979). Inertial limit on corotation. *Journal of Geophysical Research*, 84( A11), 6554– 6558. https://doi.org/10.1029/JA084iA11p06554

Janvier, M., Aulanier, G., Bommier, V., et al. (2014), Electric currents in flare ribbons: observations and three-dimensional standard model, ApJ, 788, 60.

Jasperse, J. R., Ion heating, electron acceleration, and the self-consistent parallel E-field in downward auroral current regions, Geophys. Res. Lett., 25, 3485, 1998.

Kasper, J. C., et al. (2016), Space Sci. Rev., 204, 131.

Klumpar, D. M. (1990), Near equatorial signatures of dynamic auroral processes, in Physics of Space Plasmas, pp. 265– 276, Sci. Publ., Cambridge, Mass.

Korreck, K., et al. (2109), Astrophys. J., in prep. [this issue].

Krymsky, G.F. (1977), A regular mechanism for the acceleration of charged particles on the front of a shock wave, *Dokl.~SSSR*, 234, 1306.

Lund, E.J., E. Möbius, K. A. Lynch, D. M. Klumpar, W. K. Peterson, R. E. Ergun, C. W. Carlson, (2001), On the mass dependence of transverse ion acceleration by broad-band extremely low frequency waves, Physics and Chemistry of the Earth, Part C: Solar, Terrestrial & Planetary Science, 26, 161-163 https://doi.org/10.1016/S1464-1917(00)00102-1

Lynch, K. A., J. Bonnell, C. Carlson, W. Peria, (2002), Return-current-region aurora: E-parallel,j-z, particle energization, and BBELF wave activity, J. Geophys. Res.,107, 1115, 10.1029/2001JA900134.

Marklund, G. T., et al. (2001), Temporal evolution of the electric field accelerating electrons away from the auroral ionosphere, Nature, 414, 724–727, doi:10.1038/414724a.

McComas, D. J., et al. (2016), Space Sci. Rev., 204, 187–256.

McComas, D. J., E. R. Christian, C. M. S. Cohen, et al. (2019), in press, Nature.

Melrose, D. B. (2007), Acceleration Mechanisms, arXiv:0902.1803 [astro-ph.SR]

Mitchell, D. G., et al. (2009), Ion conics and electron beams associated with auroral processes on Saturn, J. Geophys. Res., 114, A02212, doi:10.1029/2008JA013621.

Moebius, E. et al. (1999), Variation of Ionic Charge States Between Solar Energetic Particle Events as Observed With ACE SEPICA, Proc. 26th Int. Cosmic Ray Conf., 1999ICRC....6...87M

Popecki, M. A., et al. (2002), Ionic charge state measurements in solar energetic particle events, Adv. Space Res. Vol. 30, No. 1, pp. 3343,2

Reames, D.V. (1999), Particle acceleration at the Sun and in the heliosphere, *Space Sci. Rev.*, 90, 413.

Reames, D.V. (2015), What Are the Sources of Solar Energetic Particles? Element Abundances and Source Plasma Temperatures, *Space Sci Rev* 194:303—327

Reid, H. A. S., and H. Ratcliffe (2014), A Review of Solar Type III Radio Bursts, *Res. Astron. Astrophys.* 14 773

Roth, I and M. Temerin (1997), Enrichment of 3He and heavy ions in impulsive solar flares, *Astrophys. J.*, 477:940È957

Rouillard, A. P., et al. (2019), Astrophys. J., in prep. [this issue]

Schwadron, N. A., et al. (2015), Particle Acceleration at Low Coronal Compression Regions and Shocks, *Astrophys. J.*, 810, article id. 97.

Temerin, M. and I. Roth (1992), The production of He3 and heavy ion enrichment in He3-rich flares by electromagnetic hydrogen cyclotron waves, Astrophys. J. Lett. 391, L105

Vourlidas, A., and D. F. Webb (2018), Streamer-blowout Coronal Mass Ejections: Their Properties and Relation to the Coronal Magnetic Field Structure, *ApJ* 861 103

Zank, G. P., Rice, W. K. M., and Wu, C. C. (2000), Particle acceleration and coronal mass ejection driven shocks: A theoretical model, *J. Geophys. Res.*, 105( A11), 25079– 25095, doi:10.1029/1999JA000455.

Zharkova, V. V., et al. (2011), Recent Advances in Understanding Particle Acceleration Processes in Solar Flares, Space Sci Rev 159:357—420, DOI 10.1007/s11214-011-9803-y